\documentstyle[12pt]{article}
\begin{document}
\begin{titlepage}
\begin{flushright}
                Preprint IFT UWr 933/99\\
                February 1999 \\
                hep-th/9902109
\end{flushright}
\bigskip
\bigskip

\begin{center}
{\Large \bf
Light-cone formulation \\
and \\
spin spectrum\\
of\\
non-critical fermionic string\\}
\end{center}
\bigskip
\bigskip

\begin{center}
        {\bf Marcin Daszkiewicz\footnote{
                Institute of Theoretical Physics, University of Wroc{\l}aw,
                pl. Maxa Borna 9, 50-204 Wroc{\l}aw, Poland; \\
                E-mail: marcin@ift.uni.wroc.pl}}\bigskip\\
        Institute of Theoretical Physics, University of Wroc{\l}aw,
        Wroc{\l}aw, Poland \bigskip\\
        {\bf Zbigniew Hasiewicz\footnote{
                Institute of Mathematics, University in Bia{\l}ystok,
                ul.Akademicka 2, 15-267 Bia{\l}ystok, Poland;\\
                E-mail: zhas@alpha.fuwb.edu.pl}}\bigskip\\
        Institute of Mathematics, University in Bia{\l}ystok,
        Bia{\l}ystok, Poland \bigskip\\
        {\bf Zbigniew Jask\'{o}lski\footnote{
                Institute of Theoretical Physics, University of Wroc{\l}aw,
                pl. Maxa Borna 9, 50-204 Wroc{\l}aw, Poland; \\
                E-mail: jaskolsk@ift.uni.wroc.pl}} \bigskip\\
        Institute of Theoretical Physics, University of Wroc{\l}aw,
        Wroc{\l}aw, Poland\\
\end{center}
\bigskip \bigskip
\bigskip\bigskip

\begin{abstract}
A free fermionic string quantum model is constructed
directly in the light-cone variables in the range of
dimensions $1<d<10$.
It is shown that after the GSO projection
this model is equivalent
to the fermionic massive string and to the
non-critical Rammond-Neveu-Schwarz string.
The spin spectrum of the model is analysed.
For $d=4$ the character generating functions
is obtained and the particle content of first few levels is
numerically calculated.
\end{abstract} 
\thispagestyle{empty}
\end{titlepage}

\section{Introduction}

It was recently shown that the covariant quantization
of the Rammond-Neveu-Schwarz  string
modified by adding the supersymmetric Liouville sector
with vanishing cosmological constants leads in 
the dimensions $1<d<10$ to a family of tachyon
free unitary models \cite{hasiewicz99}.
For every  member of this family the Neveu-Schwarz
sector does not contain any massless states which justifies
the name -- {\it fermionic massive
string} -- introduced in \cite{hasiewicz99}.
One of  the quantum models
characterised by the largest subspace of null states is equivalent to
the non-critical RNS spinning string truncated in the 
Neveu-Schwarz sector to a tachyon free subspace of 
the fermion parity operator. It is called the {\it critical 
fermionic massive string}. The aim of the present 
paper is to develop  the light-cone formulation of this model. 
Our motivation is twofold. First of all the 
light-cone formulation seems indispensable for analysing 
the splitting-joining interaction. Secondly,
it can be used to calculate
the particle content of the model.

The paper is organised as follows. In Section 2 
we define a quantum free string model directly in the 
light-cone variables. We shall call it
the {\it fermionic non-critical light-cone string} or simply
the light-cone string.  In this model
the longitudinal degrees of freedom are described by
a background charge Fock space realisation \cite{coulombgas}
of the superconformal Verma module with the central
charge $\hat{c}=10-d$,
and the highest weight $h={1\over 2}$. A similar 
construction motivated by the Liouville theory was 
first discussed by Marnelius in the context of the non-
critical Polyakov fermionic string \cite{marnelius83}.
The GSO projection is
introduced as a  projection on a suitably chosen eigenspace
of  the  (world-sheet) fermion parity operator.

In Section 3 we show that after the GSO projection the
fermionic non-critical light-cone string is equivalent 
to the critical massive string and therefore to the
suitably projected Rammond-Neveu-Schwarz non-critical string.

In Section 4 the spectrum of the  light-cone  string
is analysed. For $d=4$ the  expansion of the character 
generating function in terms of irreducible characters 
is derived. It is illustrated by numerical calculations 
of spin content of first few levels.
The spectrum of the GSO projected tachyon free model is also
calculated.
The corresponding results for the closed non-critical light-cone
string are presented in Section 5.
An interesting feature of the model is that
the closed string spectrum does
not contain space-time fermions.

 The spectra of the open and the closed critical massive
strings derived in this paper exclude the
fundamental string  interpretation of the model.
It might be however a good candidate for an effective low
energy description of strong interactions.
It shares two important features of the
critical fermionic string - the light-cone formulation
and the absence of tachyons. Whether it is enough for a
consistent interaction is an interesting open problem.

\section{Fermionic non-critical light-cone string}

Let us fix a light-cone frame
$\{e_{\pm}, e_1,\dots,e_{d-2}\}$
in $d$-dimensional Minkowski space  
normalised by
$e^2_\pm =0$, $e_+\cdot e_- =-1$, and $e_i\cdot e_j =\delta_{ij}$ for
$i,j=1,\dots,d-2$. We shall use the following  notation for the
light-cone components of a vector $V$
$$
V^\pm = e_{\pm}\cdot V\;\;,\;\;V^i\;=\;e_i\cdot V\;\;\;,
\;\;\;\overline{V}\;=\;V^ie_i\;\;\;.
$$
The fermionic non-critical light-cone string is defined as a
representation of the algebra
$$
{[a_0^i,q_0^j]}
= - i \delta^{ij}\;\;\;
,\;\;\;{[a_0^+,q_0^-]}\;=\;i\;\;\;
,\;\;\;[c_0,q_0^L]\;=\;-i\;\;\;,
$$
\begin{equation}
\label{ccr}
\begin{array}{cllcll}
{[a^i_m,a^j_n]}
&=\; m \delta^{ij} \delta_{m,-n}\;\;\;
&,\;\;\;
&[c_m,c_n]
&=\; m \delta_{m,-n}\;\;\;&,\\[6pt]
\{b^i_r,b^j_s \}
&=\; \delta^{ij} \delta_{r,-s}\;\;\;
&,\;\;\;
&\{d_r,d_s \}
&=\; \delta_{r,-s}\;\;\;&,
\end{array}
\end{equation}
supplemented by  the conjugation properties
$$
\begin{array}{lllllllll}
\left(a^i_0\right)^\dagger\!\!\!&=\; a^i_0\!\!\!&,\;\;
\left(q_0^i\right)^\dagger\!\!\!&=\;q_0^i\!\!\!&,\;\;
\left(c_0\right)^\dagger\!\!\!&=\; c_0\!\!\!&,\;\;
\left(q_0^L\right)^\dagger\!\!\!&=\;q_0^L\!\!\!&,\\[6pt]
\left(a^+_0\right)^\dagger\!\!\!&=\; a^+_0\!\!\!&,\;\;
\left(q_0^-\right)^\dagger\!\!\!&=\;q_0^-\!\!\!&,\;\;
&
&\\[6pt]
\left( a^i_m \right)^\dagger\!\!\!& =\; a^i_{-m}\!\!\!&,\;\;
\left( b^i_r \right)^\dagger\!\!\!&=\; b^i_{-r}\!\!\!&,\;\;
\left( c_m \right)^\dagger\!\!\!& =\; c_{-m}\!\!\!&,\;\;
\left( d_r \right)^\dagger\!\!\!& =\; d_{-r}\!\!\!&,
\end{array}
$$
where $m,n\in Z\!\!\!Z;r,s \in Z\!\!\!Z + {\textstyle{\epsilon\over 2}}$.
The operators $P^+=\sqrt{\alpha}a^+_0, P^i=\sqrt{\alpha}a^i_0$, and
$x^-={1\over \sqrt{\alpha}}q^-_0, x^i={1\over \sqrt{\alpha}}q^i_0$
are interpreted as components of the total momentum of the string,
and the barycentric coordinates, respectively.

Let us denote by $F_\epsilon(p^+,\overline{p})$  the
Fock space generated by the algebra of non-zero modes (with negative labels)
out of the unique vacuum state $\Omega_\epsilon$
satisfying
$$
P^i\,\Omega_\epsilon = p^i \,\Omega_\epsilon\;\;\;,\;\;\;
P^+\,\Omega_\epsilon = p^+ \,\Omega_\epsilon\;\;\;,\;\;\;
c_0\,\Omega_\epsilon = \lambda\,\Omega_\epsilon\;\;\;.
$$
The space of states is a direct integral of Hilbert spaces
over the  spectrum of momentum operators
$$
 H_\epsilon
=  \int {dp_{+}\over |p_+|} d^{d-2}\overline{p}\;
H_\epsilon(p^+,\overline{p})\;\;\;.
$$
In the Neveu-Schwarz sector ($\epsilon = 1$)
$$
H_1(p^+,\overline{p})=F_1(p^+,\overline{p})\;\;\;.
$$
In the Rammond sector ($\epsilon = 0$) the fermionic zero modes
$b^i_0,d_0$
form the real Euclidean Clifford algebra ${\cal C}(d-1,0)$.
If one requires a well defined fermion parity
operator  the zero mode sector of $H_0(p^+,\overline{p})$ must carry a
representation of the Clifford algebra ${\cal C}(d,0)$.
We assume that this sector is described  by
an irreducible representation $D(d)$ of the complexified Clifford
algebra ${\cal C}^C(d) = {\cal C}(d,0)\otimes C\!\!\!\!I$, and
$$
H_0(p^+,\overline{p})=F_0(p^+,\overline{p})\otimes D(d)\;\;\;.
$$
The representation of the algebra (\ref{ccr})
on $H_0(p^+,\overline{p})$ is given by
$$
\begin{array}{rcllrclll}
a^i_m &=& {\tilde a}^i_m\otimes 1\;&,\;\;\;&
c_m&=& {\tilde c}_m \otimes 1\;&,\;\; m\neq 0\;&,\\[6pt]
b^i_r &=&  {\tilde b}^i_r\otimes \Gamma^F\;&,\;\;\;&
d_r&=& {\tilde d}_r \otimes\Gamma^F\;&,\;\; r\neq 0\;&,\\[6pt]
b^i_0 &=& 1\otimes {\textstyle {1\over \sqrt{2}}}\Gamma^i  \;&,\;\;\;&
d_0 &=& 1\otimes{\textstyle {1\over \sqrt{2}}}\Gamma^L \;&,\;\;
\end{array}
$$
where ${\tilde a}^i_m,{\tilde c}_m,{\tilde b}^i_r,{\tilde d}_r$
denote the operators on $F_0(p^+,\overline{p})$
representing  the non-zero bosonic and
fermionic modes, and $\Gamma^1,...,\Gamma^{d-2},\Gamma^L,\Gamma^F$
are the gamma matrices of the $D(d)$ representation.

In order to construct generators of a unitary representation of
the Poincare group we introduce the operators
\begin{eqnarray*}
  L_m  & = & {\textstyle{1\over 2}}\!\! \sum\limits_{n \in Z\!\!\!Z}
  :\overline{a}_{-n}\cdot \overline{a}_{n+m}: \;+\;
{\textstyle{1\over 2}} \!\!\!
 \sum\limits_{r \in Z\!\!\!Z + {\epsilon\over 2} }
   r \,:\overline{b}_{-r} \cdot \overline{b}_{r+m}: \;+\;
   (1-\epsilon) {\textstyle{d+1\over 16}}\delta_{m,0}\nonumber\\
& + &
{\textstyle{1\over 2}}\!\! \sum\limits_{n \in Z\!\!\!Z}
: c_{-n}c_{n+m}:\; +\;
{\textstyle{1\over 2}} \!\!\!\sum\limits_{r \in Z\!\!\!Z +
{\epsilon \over 2} }
 r\, :d_{-r} d_{r+m}:\;+\;
 2i\sqrt{\beta}m c_m \;+ \;2\beta\delta_{m0}\;\;\;, \\
G_r   & = &
 \sum\limits_{n \in Z\!\!\!Z}
\overline{a}_{-n}\cdot \overline{b}_{n+r}\; +\;
 \sum\limits_{n \in Z\!\!\!Z}
c_{-n}d_{n+r}\; + \;4i\sqrt{\beta} r d_r \;\;\;,\nonumber
\end{eqnarray*}
forming an $N=1$ superconformal algebra with the central charge
$\hat{c} = d-1+32\beta$
\begin{eqnarray*}
   \left[ L_m, L_n \right]   & = & (m-n)L_{m+n}
   + {\textstyle\frac{1}{8}}(d-1+32\beta)(m^3-m)\delta_{m,-n}
                             \;\;\;,\nonumber   \\[6pt]
   \left[ L_m, G_r \right]   & = & ({\textstyle\frac{1}{2}}m - r)G_{m+r}
   \;\;\;,\\[6pt]
   \left\{ G_r, G_s \right\} & = & 2L_{r+s}
                + {\textstyle{1\over 2}}(d-1+32\beta)(r^2-
{\textstyle{1\over 4}})\delta_{r,-s}\;\;\;. \nonumber
\end{eqnarray*}
The generators of the translations in the longitudinal and the
transverse directions are given by
the operators $P^+$ and $P^i$, respectively.
The generator of the translation in the $x^+$-direction  is
defined by
$$
{P}^- =
{\frac{\alpha}{P^+}} (L_0 - \alpha_0) \;\;\;.
$$
The $x^+$ coordinate is regarded as an evolution parameter
and $P^-$ plays the role of the string Hamiltonian.
The generators of the Lorentz group are defined
by
\begin{eqnarray*}
M^{ij}& = &  x^i P^j - x^j P^i -
i\sum\limits_{n>0}
{{1\over n}} ( a^i_{-n} a^j_{n} - a^j_{-n} a^i_{n}
)\;\;\;\nonumber\\
&&\hspace{10pt}
+\; (1- \epsilon) i b^i_0 b^j_0
- i\sum\limits_{r>0}
( b^i_{-r} b^j_{r} - b^j_{-r} b^i_{r}) \;\;\;,  \nonumber\\
M^{i+}&=& x^iP^+\;\;\;,\\
M^{+-}& =& {\textstyle{1\over 2}} (P^+x^- + x^-P^+)\;\;\;,\\
M^{i-}& = &  {\textstyle{1\over 2}} (P^-x^i + x^iP^-) - x^- P^i -
{i\over a^+_0}\sum\limits_{n>0}
{{1\over n}} ( a^i_{-n} L_{n} - L_{-n} a^i_{n}
)\;\;\;\nonumber\\
&&\hspace{10pt}
+\; (1- \epsilon) {i\over a^+_0} b^i_0 G_0- {i\over a^+_0}\sum\limits_{r>0}
( b^i_{-r} G_{r} - G_{-r} b^i_{r}
)  \;\;\;,
\end{eqnarray*}
The algebra of the generators $P^+,P^-,P^i$, $M^{+-},M^{i+},M^{i-},
M^{ij}$ closes to the Lie algebra of the Poincare group
up to some anomalous terms. They appear only in the commutators
\begin{eqnarray*}
\left[\; {M}^{i-}\;,\;{M}^{j-} \right]& =&
\frac{1}{8{a_0^+}^2} \sum_{n>0} (\Delta n - \widetilde{\Delta}
{1\over n} )
(a_{-n}^ia_{n}^j - a_{-n}^ja_{n}^i) \\
&+ &
\frac{1}{2{a_0^+}^2} \sum_{r>0}
(\Delta  - \widetilde{\Delta} {1\over r^2} )
(b_{-r}^ib_{r}^j - b_{-r}^jb_{r}^i) \;\;\;,
\end{eqnarray*}
where $\Delta = d-9+ 32\beta,\;
\widetilde{\Delta} = 16\alpha_0 -d+1-32\beta $,
and vanish if and only if $\beta= {1\over 32}(9-d)$,
and $\alpha_0={1\over 2}$.
The first condition implies that the operators
$P^-, M^{i-}$
are self-adjoint only in the range $2\leq d\leq9$.
The second leads to the following expression
for the mass square operator
\begin{equation}
\label{mass-shell}
M^2_\epsilon = 2\alpha \textstyle ( R_\epsilon +
{\lambda^2\over 2} - \epsilon{d-1\over 16})
\;\;\;,
\end{equation}
where $R_\epsilon  =  \sum\limits_{m>0}
  (\overline{a}_{-m}\cdot \overline{a}_{m} +
  c_{-m}c_m ) + \sum\limits_{r>0}
   r (\,\overline{b}_{-r} \cdot \overline{b}_{r+m} +d_{-r} d_{r+m})$
is the level operator.
Note that in the covariant  massive string model
the eigenvalue $\lambda$ of the bosonic Liouville zero mode
$c_0$ is restricted by the constraint $c_0=0$.
In the present construction it is regarded as a free
real parameter.

It follows from (\ref{mass-shell}) that for $\lambda^2$ small enough
the ground states in the Neveu-Schwarz sector are tachyonic.
One can try to solve this problem by introducing
the GSO projection \cite{gso}.
Let  $\tilde{F_\epsilon} = \sum_{r>0}
\overline{\tilde{b}}_{-r} \cdot \overline{\tilde{b}}_{r}
+ \sum_{r>0}\tilde{d}_{-r}\tilde{d}_{r}$ be the fermion number
operator on $F_\epsilon(p^+,\overline{p})$.
We introduce the fermion parity operators on
the total Hilbert space
$H=H_0 \oplus H_1$ :
$$
 (-1)^F =  (-1)^{\tilde{F_0}} \otimes \Gamma^F
 \oplus (-1)^{\tilde{F}_1+1}\;\;\;.
$$
The GSO projection is  defined as the projection on the
$+1$ eigenspace of $(-1)^F$.
In the case of even dimensions
there exists another operator
$\Theta$ with all the properties of the
fermion parity operator, and anticommuting with  $(-1)^F$.
One can show that the GSO
projections with respect to $\Theta$, and $(-1)^F$ lead to
equivalent models.

\section{Equivalence to other models}

In this section we shall show that the light-cone string
is equivalent to
the critical fermionic massive string recently introduced
in \cite{hasiewicz99}.
In the covariantly quantized fermionic massive string
the conditions for physical states
can be solved in terms of the transversal $A^i_{m}, B^i_r$, the
super-Liouville $C_m, D_r$, and the "shifted" longitudinal
$A^{\scriptstyle L}_m, B^{\scriptstyle L}_r$.
For details concerning the DDF construction
and the notation used in this section we refer to \cite{hasiewicz99}.

The critical fermionic massive string corresponds to
a special choice of the  parameters
$\beta={9-d\over 32}, m_0^2 =0$. In this case all
states containing the "shifted" longitudinal
excitations are null. The space of physical states can
be identified with all states generated by  the
transverse and the super-Liouville DDF operators
$A^i_{m}, B^i_r$, $C_m, D_r$. They have the same
(anti)commutation relations
and the conjugation properties,
as the light-cone excitations $a^i_{m}, b^i_r$, $c_m, d_r$.
Also the continuous spectra of the 
bosonic zero modes in both models are identical.
In the critical massive string
the representation of the transverse, the Liouville,
and the fermion parity gamma matrices on the
on-mass-shell physical states
coincides with the representation $D(d)$.
The only difference is that in the covariant
model one gets a neutral, while in the
light-cone string a positive definite scalar product.
This discrepancy is not essential - the subspaces
with definite products in the covariant model
are eigenspaces of the the fermion parity operator \cite{hasiewicz99}.
Note that the neutral product of the covariant model
is a consequence of the assumption that
the zero and the non-zero fermionic modes
have the same conjugation properties.

One way to show the equivalence of
the Poincare group representations
is to calculate the commutators of the Poincare generators with
the DDF operators. This calculations can be
facilitated by the technique of the leading terms \cite{goddard72}.
It is based on the observation that the DDF operators
expressed in terms of elementary excitations are
uniquely determined by they leading terms i.e.
parts of such expressions which do not contain any
$a^+_m, b^+_r; m\neq 0$ excitations.
The most tedious calculations are involved in the
commutators:
\begin{eqnarray*}
\left[M^{i-},A_m^j\right]
& = &
{\textstyle{m\over a^+_0}} {\rm e}^{im{\scriptstyle{q_0^+\over a^+_0}}}
\!\!\left(
{\textstyle\frac{ q_0^+}{a_0^+}}a^i_0A_m^j
- q^i_0 A_m^j
-{\textstyle {\frac{i}{m}\delta^{ij}}}
(A_m^{\scriptscriptstyle L} - {\cal L}_m )
  -
i(1-\epsilon) B_0^iB^j_m \right.\\
& + &\left.
i\!\sum_{n>0} {\textstyle\frac{1}{n}}( A_{n+m}^j A_{-n}^i
\!-\! A_{-n+m}^j A_n^i )
+i\!\sum_{r>0} ( B_{r+m}^j B_{-r}^i\! +\! B_{-r+m}^j B_r^i )
\right) \\
\left[M^{i -},B_r^j\right]
& = &
{\textstyle{1\over a^+_0}} {\rm e}^{ir{\scriptstyle{q_0^+\over a^+_0}}}
\left(
r{\textstyle\frac{ q_0^+}{a_0^+}}a^i_0B_r^j
- rq^i_0 B_r^j
-i\delta^{ij}
(B_r^{\scriptscriptstyle L} - {\cal G}_r )
 -i(1-\epsilon) B_0^iA^j_r \right. \\
& + &
i\!\sum_{n>1}\left(
{\textstyle\frac{r}{n}}(  A_{-n}^i B_{n+r}^j \!-\!
 A_n^i B_{r-n}^j)
+ {\textstyle\frac{1}{2}}( A_{-n}^i B_{n+r}^j \!+
\! A_n^i B_{r-n}^j) \right)\\
& - &   \left.
i\sum_{s>0} (A_{r+s}^j B_{-r}^i\! +\!
A_{r-s}^j B_r^i)
 \right)\\
\left[M^{i-},C_m\right]
& = &
{\textstyle{m\over a^+_0}} {\rm e}^{im{\scriptstyle{q_0^+\over a^+_0}}}
\!\!\left(
{\textstyle\frac{ q_0^+}{a_0^+}}a^i_0C_m
- q^i_0 C_m
-2\sqrt{\beta} A_m^i
- i(1-\epsilon) B_0^iD_m \right.\\
& + &\left.
i\!\sum_{n>0} {\textstyle\frac{1}{n}}( C_{n+m} A_{-n}^i
\!-\! C_{-n+m} A_n^i )
+i\!\sum_{r>0} ( D_{r+m}B_{-r}^i\! +\! D_{-r+m} B_r^i )
\right) \\
\left[M^{i -},D_r\right]
& = &
{\textstyle{1\over a^+_0}} {\rm e}^{ir{\scriptstyle{q_0^+\over a^+_0}}}
\left(
r{\textstyle\frac{ q_0^+}{a_0^+}}a^i_0D_r
- rq^i_0 D_r
+4\sqrt{\beta}r B_r^i
 -i(1-\epsilon) B_0^iC_r \right. \\
& + &
i\!\sum_{n>1}\left(
{\textstyle\frac{r}{n}}(  A_{-n}^i D_{n+r}-
 A_n^i D_{r-n})
+ {\textstyle\frac{1}{2}}( A_{-n}^i D_{n+r} +
 A_n^i D_{r-n}) \right)\\
& - &   \left.
i\sum_{s>0} (C_{r+s} B_{-r}^i +
C_{r-s} B_r^i)
 \right)\;\;\;.
\end{eqnarray*}
Setting in these formulae the evolution parameter
$q_0^+= {x^+0\over \sqrt{\alpha}}=0$,
and neglecting the "shifted"
longitudinal DDF operators $A^L_m, B^L_r$ one reproduces
the corresponding light-cone commutators.
One can easily check
that this is also true for all the other generators.

We have shown that the GSO projected fermionic light-cone string
is isomorphic to the GSO projected critical massive string.
The latter model is  equivalent
to the  non-critical Rammond-Neveu-Schwarz string truncated
in the Neveu-Schwartz sector to the tachion free eigenspace of the
fermion parity operator \cite{hasiewicz99}. One thus has three
equivalent descriptions of the same fermionic non-critical string
model.

\section{Spin spectrum}

The problem of  the spin spectrum is to decompose
the unitary representation
of the Poincare group on the Hilbert space of string
into irreducible
representation.
It follows from formula (\ref{mass-shell}) that
the decomposition of
$H_\epsilon$ into representations of a fixed
mass coincides with the level structure
$$
H_\epsilon =
\bigoplus_{N \geq 0}
\int {dp_{+}\over |p_+|} d^{d-2}\overline{p}
\;H_\epsilon^{N}(p^+,\overline{p})\;\;\;,
\;\;\;\;\;
R_\epsilon \;H_\epsilon^{N}(p^+,\overline{p})\;
=\; NH_\epsilon^{N}(p^+,\overline{p})\;\;\;.
$$
For $\lambda^2$ in the range $0 \leq \lambda^2 < {d-1\over 8}$
the lowest level subspace $H_1^{0}$ in the Neveu-Schwarz
sector carries an irreducible tachyonic representation.
For $\lambda^2 = {d-1\over 8}$, $H_1^{0}$ is a massless,
and for $\lambda^2 > {d-1\over 8}$, a massive scalar
representation.

In the Rammond sector the
0-level subspaces $H_0^{0}(p^+,\overline{p})$
 are by construction isomorphic with
the irreducible representation $D(d)$ of the complex Clifford algebra
${\cal C}^C(d)$. For a massive momentum ($\lambda^2>0$)
the representation
of the little group ${\rm Spin}(d-1)$ on $D(d)$
is a direct sum of two isomorphic
fundamental irreducible representations
$S(d-1)$ of ${\rm Spin}(d-1)$.

In the massless case ($\lambda^2 =0$) the maximal compact subgroup
of the little group is ${\rm Spin}(d-2)$.
In the odd dimensions $D(d)$ is a direct sum of two
fundamental irreducible representations of ${\rm Spin}(d-2)$ while
in the  even dimensions it is a direct sum of four such representations.
In particular for $d=4$ and $\lambda^2 =0$ the zero level in the
Rammond sector contains two pairs of the left, and the right Weyl spinors.

Since all higher levels  are massive, the spaces
$H^N_\epsilon(p^+,\overline{p})$ should be decomposed into
 irreducible representations of the little group
${\rm Spin}(d-1)$.
For every momentum $p$ with $m^2 =
\alpha(2N +\lambda^2 -{\varepsilon}\frac{d-1}{8}), N>0$,
one can choose a light-cone frame such that
$p^+ = {\sqrt{\alpha}}, \overline{p} =0$,
and the little group  is generated by
$$
{\cal G}^j = M^{j-} - {m^2 \over 2\alpha} M^{j+}
\;\;\;\;,\;\;\;\;M^{ji} \;\;\;\;.
$$
We shall use the method developed in the case of
the bosonic light-cone string
\cite{daszkiewicz98}. It relies on the observation that,
as far as  the character generating function is concerned,
the vector representation of ${\rm Spin}(d-2)$ formed by the transverse
excitation can be extended
to a vector representation of ${\rm Spin}(d-1)$
by means of the Liouville  excitations.
The only novelty is that in the present case  we have two
vector representations $V_{-m}^{\rm B}$ and $V_{-r}^{\rm F}$
spanned by the creation operators
$$
   a_{-m}^a  = \left\{
   \begin{array}{rl}
   \kappa_m\;c_{-m} & \;
   a=0\\
   a _{-m}^a &\; 1\leq a\leq d-2
  \end{array} \right. \;,\;
   b_{-r}^a =  \left\{
   \begin{array}{rl}
   \kappa_r\;d_{-r}
   &      a=0\\
   b _{-r}^a &  1\leq a\leq d-2
  \end{array} \right.  \;.
$$
The normalisation constants
$\kappa_m=\frac{\lambda - 2i\sqrt{\beta}k}{\sqrt{\lambda^2 +4\beta k^2}}$,
$\kappa_r=\frac{\lambda - 4i\sqrt{\beta}r}{\sqrt{\lambda^2 + 16\beta r^2}}$
are chosen in order to obtain the canonical
antisymmetric matrix generators $D^{(i)}$
of ${\rm Spin}(d-1)$ vector representation:
\begin{eqnarray*}
\left[\; {\cal G}^i \;,\; A^a_{-m} \right]\;&=&\;
   i {\sqrt{\lambda^2 + 4m^2\beta}}\; {D^{(i) \, a}}_b A^b_{-m} +
    \dots\;\;\;,  \\
\left[\; {\cal G}^i\;,\;B_{-r}^a\;\right]\;&=&\;
i\sqrt{\lambda^2 +16r^2\beta}\; {D^{(i)\,a}}_bB_{-r}^b\;\, +
\dots \;\;\;.
\end{eqnarray*}
The dots in the formulae above denote all
terms of higher order in the excitation operators.
Such terms do not contribute to the character functions.

The subspace $H^N_\epsilon(\sqrt{\alpha},0)$
decomposes into a direct sum
of tensor products of the symmetric tensor powers of
$V_{-m}^{\rm B}$, the  antisymmetric tensor powers of $V_{-r}^{\rm F}$,
and $D(d)$. Then using the method of  \cite{daszkiewicz98} one can write
the character of the ${\rm Spin}(d-1)$ representation on
$H^N_\epsilon(\sqrt{\alpha},0)$  as
$$
\chi^N_\epsilon =
 \sum_{N_{\rm B}+N_{\rm F}=N}\,
 \sum_{p_{\rm B}\in P(N_{\rm B})}\,
 \sum_{p_{\rm F}\in P(N_{\rm F}) } \,
 \prod_{m_k \in p_{\rm B}}           \,
 \prod_{m_r \in p_{\rm F}}
 \chi_{S}^{m_k}\chi_{A}^{m_r}\chi^0_\epsilon \;,
$$
where the sum runs over all partitions
$p_{\rm B}\!=\!\{m_k\}$,
$p_{\rm F}\!=\!\{m_r\}$
of the  bosonic $N_{\rm B}$, and the  fermionic
$N_{\rm F}$ level number. The symbols
$\chi_{S}^{m_k}$, and  $\chi_{A}^{m_r}$ stand
for the characters of the $m_k$-th symmetric, and the $m_r$-th
antisymmetric tensor power of the vector
representation of ${\rm Spin}(d-1)$, respectively. Finally,
$\chi^0_\epsilon$ is given by
\begin{equation}
\label{doubling}
\chi^0_0= 2 \chi_{S(d-1)}\;\;\;,\;\;\;\chi^0_1=1\;\;\;,
\end{equation}
where $\chi_{S(d-1)}$ is the character
of the fundamental irreducible representation of ${\rm Spin}(d-1)$.
Using the formulae for characters of tensor products \cite{weyl}
one gets the character  generating function
$$
\chi_\epsilon(t,g) = \sum_{N\ge 0}t^N \chi^N (g)\,=
 \prod\limits_{ k \in I\!\!I}
 { 1\over
{\rm det}(1-t^k{\cal D}_{\rm v}(g))}
\prod\limits_{ (1+\epsilon)r \in I\!\!I}
{\rm det}(1+t^r{\cal D}_{\rm v}(g))
\, \chi^0_\epsilon(g)
\;\;\;,
$$
where ${\cal D}_{\rm v}$ denotes the vector representation of
${\rm Spin}(d-1)$, and $I\!\!I$ is the set of all positive integers.
The expansions of $\chi_\epsilon(t,g)$ in terms of
irreducible characters can be found using the
techniques  developed
by Curtright and Thorn \cite{curtright86}
for strings with only transverse excitations.
In the case of $d=4$ one gets:
\begin{eqnarray*}
\chi_{\epsilon} (t,\varphi )& =&
2^{1-\epsilon}t^{\frac{1}{8}(1-\varepsilon )}p^4(t)
{\pi}_{\epsilon}(t)
\!\!\!\!\!\!
\sum_{l \in I\!\!I +{\epsilon-1\over 2}}
        \chi_l(\varphi)\\
&&
\sum_{k \in I\!\!I }
(-1)^{k-1}(1-t^k)
\sum\limits_{m \in I\!\!I +{\epsilon-1\over 2}}
t^{\frac{(k(k-1)+m^2)}{2}}\;
(1-t^{m+\frac{1}{2}})(t^{k|l-m|}-t^{k(l+m+1)})
\;\;\;,
\end{eqnarray*}
where
$$
p(t)=
\prod\limits_{n \in I\!\!I }
{(1-t^n)}^{-1} \;\;\;\;\;,\;\;\;\; \pi_{\epsilon} (t) =
\prod\limits_{ (1+\epsilon)r \in I\!\!I }
(1+t^r)\;\;\;,
$$
and $\chi_l(\phi)$ is the character of the
spin $l$ irreducible representation of ${\rm Spin}(d-1)$.
For $\lambda^2=0$ the spin spectrum up to 6-th level is
presented on Fig.1.

The doubling of the spectrum in the Rammond sector
(\ref{doubling})
is related to the
presence of the fermionic zero mode in the Liouville sector.
For all dimensions in the range $1<d<10$
the GSO projection removes
this doubling without any extra conditions for the parameters
of the model. In the Neveu-Schwarz sector it simply removes
the integer levels.
The GSO truncated spectrum is presented on Fig.2.

\section{Closed string}

The closed string Hilbert space
$H_c$ can be constructed as the subspace
of the tensor product of two copies
of the open string Hilbert spaces
$
(H_0\oplus H_1)
\otimes (\widetilde{H}_0\oplus \widetilde{H}_1)
$
determined by the conditions
$$
a^i_0 = \widetilde{a}^i_0 = {P^i_c\over 2\sqrt{\alpha}}\;\;\;,
\;\;\;
a^+_0 = \widetilde{a}^+_0 = {P^+_c\over 2\sqrt{\alpha}}\;\;\;,
\;\;\;c_0 = \widetilde{c}_0 =\lambda \;\;\;,
$$
and anihillated by the twist operator
$T = (R_0 \oplus R_1)\otimes 1 - 1\otimes(\widetilde{R}_0 \oplus
\widetilde{R}_1)$.

Since the mass levels of different sectors of
the open non-critical string never coincide (\ref{mass-shell})
the mixed sectors $H_0\otimes \widetilde{H}_1$,
$H_1\otimes \widetilde{H}_0$ are excluded.
In consequence the spectrum of the closed fermionic string
does not contain space-time fermions. This is in fact
a common feature of all the covariant closed string models
corresponding to the family of non-critical
open strings considered in \cite{hasiewicz99}.

The representation of the Poincare generators are constructed
in a standard manner. In particular, the Hamiltonian $P^-$
generating the $x^+$-evolution, and the mass square
operator $M^2\epsilon$ are given by
$$
{P}_c^-=
{\frac{\alpha}{P^+}}
(L_0 + \widetilde{L}_0 -1)\;\;\;\;,\;\;\;
M^2_\epsilon  = 4\alpha (R +\widetilde{R} +
\lambda^2   - \epsilon \textstyle{{d-1 \over 8}} )\;\;\;.
$$
The character generating function can be calculated as the "diagonal" part
(i.e. all terms of the form $t^N {t'}^N$ )
of the product of two open string generating functions
$$
\chi^{(\rm closed)}_\epsilon(\,t,g) =
{\rm Diag}(\chi_\epsilon(t,g)
\chi_\epsilon(t',g))\,|_{t=t'}\;.
$$
For $d=4$, and $\lambda^2=0$
the results of the numerical calculations of first few levels,
before, and after GSO projection are presented on Fig.3, and on Fig.4,
respectively.

\section*{Acknowledgements}

The authors would like to thank Andrzej Ostrowski
for many stimulating discussions.
This work is supported by the Polish Committee of Scientific Research
(Grant Nr. PB 1337/PO3/97/12).

\newpage
\begin{picture}(400,530)(-190.00,40.00)
\special{em:linewidth 0.4pt}
\unitlength 1.10mm
\linethickness{0.4pt}
\put(-53.25,27.00){\vector(0,1){150.00}}
\multiput(-53.25,49.67)(0.00,10.00){13}{\line(-1,0){1.00}}
\put(-60.00,49.67){\makebox(0,0)[r]{\small $\frac{1}{2}$}}
\put(-60.00,59.67){\makebox(0,0)[r]{\small $1$}}
\put(-60.00,69.67){\makebox(0,0)[r]{\small $\frac{3}{2}$}}
\put(-60.00,79.67){\makebox(0,0)[r]{\small $2$}}
\put(-60.00,89.67){\makebox(0,0)[r]{\small $\frac{5}{2}$}}
\put(-60.00,99.67){\makebox(0,0)[r]{\small $3$}}
\put(-60.00,109.67){\makebox(0,0)[r]{\small $\frac{7}{2}$}}
\put(-60.00,119.67){\makebox(0,0)[r]{\small $4$}}
\put(-60.00,129.67){\makebox(0,0)[r]{\small $\frac{9}{2}$}}
\put(-60.00,139.67){\makebox(0,0)[r]{\small $5$}}
\put(-60.00,149.67){\makebox(0,0)[r]{\small $\frac{11}{2}$}}
\put(-60.00,159.67){\makebox(0,0)[r]{\small $6$}}
\put(-60.00,169.67){\makebox(0,0)[r]{\small $\frac{13}{2}$}}

\put(-70.00,39.67){\vector(1,0){140.00}}
\put(60,30.67){\makebox(0,0)[cc]{\small $m^2/\alpha$}}
\put(-52.375,175.00){\makebox(0,0)[l]{\small spin}}
\put(3.00,25.00){\makebox(0,0)[cc]{\rm Fig.1}}

\put(-50.00,130.00){\framebox(35,17)[cc]{\shortstack[l]{$D=4$\\
                                            {\circle*{2.00}}
                                            $\alpha (0)_{\rm NS}
                                            = -\frac{3}{16}$\\
                                            {\circle{2.00}}
                                            $\alpha (0)_{\rm R}
                                            = 0$ }}}

\put(-57.00,39.67){\circle*{2.00}}
\put(-47.00,59.67){\circle*{2.00}}
\put(-37.00,59.67){\circle*{2.00}}

\put(-53.25,49.67){\circle{2.00}}
\put(-33.25,49.67){\circle{2.00}}
\put(-33.25,69.67){\circle{2.00}}
\put(-13.25,49.67){\circle{2.00}}
\put(-13.25,69.67){\circle{2.00}}
\put(-13.25,89.67){\circle{2.00}}
\put(6.75,49.67){\circle{2.00}}
\put(6.75,69.67){\circle{2.00}}
\put(6.75,89.67){\circle{2.00}}
\put(6.75,109.67){\circle{2.00}}
\put(26.75,49.67){\circle{2.00}}
\put(26.75,69.67){\circle{2.00}}
\put(26.75,89.67){\circle{2.00}}
\put(26.75,109.67){\circle{2.00}}
\put(26.75,129.67){\circle{2.00}}
\put(46.75,49.67){\circle{2.00}}
\put(46.75,69.67){\circle{2.00}}
\put(46.75,89.67){\circle{2.00}}
\put(46.75,109.67){\circle{2.00}}
\put(46.75,129.67){\circle{2.00}}
\put(46.75,149.67){\circle{2.00}}
\put(66.75,49.67){\circle{2.00}}
\put(66.75,69.67){\circle{2.00}}
\put(66.75,89.67){\circle{2.00}}
\put(66.75,109.67){\circle{2.00}}
\put(66.75,129.67){\circle{2.00}}
\put(66.75,149.67){\circle{2.00}}
\put(66.75,169.67){\circle{2.00}}

\put(-27.00,39.67){\circle*{2.00}}
\put(-27.00,59.67){\circle*{2.00}}
\put(-27.00,79.67){\circle*{2.00}}
\put(-17.00,39.67){\circle*{2.00}}
\put(-17.00,79.67){\circle*{2.00}}
\put(-17.00,59.67){\circle*{2.00}}
\put(-7.00,39.67){\circle*{2.00}}
\put(-7.00,79.67){\circle*{2.00}}
\put(-7.00,59.67){\circle*{2.00}}
\put(-7.00,99.67){\circle*{2.00}}

\put(-57,43.67){\makebox(0,0)[cc]{\small 1}}
\put(-47,63.67){\makebox(0,0)[cc]{\small 1}}
\put(-37,63.67){\makebox(0,0)[cc]{\small 2}}
\put(-27,83.67){\makebox(0,0)[cc]{\small 1}}
\put(-27,63.67){\makebox(0,0)[cc]{\small 2}}
\put(-27,43.67){\makebox(0,0)[cc]{\small 2}}
\put(-17,63.67){\makebox(0,0)[cc]{\small 3}}
\put(-17,83.67){\makebox(0,0)[cc]{\small 3}}
\put(-17,43.67){\makebox(0,0)[cc]{\small 3}}
\put(-7,103.67){\makebox(0,0)[cc]{\small 1}}
\put(-7,63.67){\makebox(0,0)[cc]{\small 7}}
\put(-7,83.67){\makebox(0,0)[cc]{\small 4}}
\put(-7,43.67){\makebox(0,0)[cc]{\small 3}}

\put(-51.20,53.67){\makebox(0,0)[cc]{\small 2}}
\put(-33.20,73.67){\makebox(0,0)[cc]{\small 4}}
\put(-33.20,53.67){\makebox(0,0)[cc]{\small 4}}
\put(-13.20,93.67){\makebox(0,0)[cc]{\small 4}}
\put(-13.20,73.67){\makebox(0,0)[cc]{\small 12}}
\put(-13.20,53.67){\makebox(0,0)[cc]{\small 12}}
\put(6.80,113.67){\makebox(0,0)[cc]{\small 4}}
\put(6.80,93.67){\makebox(0,0)[cc]{\small 16}}
\put(6.80,73.67){\makebox(0,0)[cc]{\small 32}}
\put(6.80,53.67){\makebox(0,0)[cc]{\small 32}}
\put(26.80,133.67){\makebox(0,0)[cc]{\small 4}}
\put(26.80,113.67){\makebox(0,0)[cc]{\small 16}}
\put(26.80,93.67){\makebox(0,0)[cc]{\small 48}}
\put(26.80,73.67){\makebox(0,0)[cc]{\small 84}}
\put(26.80,53.67){\makebox(0,0)[cc]{\small 72}}
\put(46.80,153.67){\makebox(0,0)[cc]{\small 8}}
\put(46.80,133.67){\makebox(0,0)[cc]{\small 16}}
\put(46.80,113.67){\makebox(0,0)[cc]{\small 52}}
\put(46.80,93.67){\makebox(0,0)[cc]{\small 128}}
\put(46.80,73.67){\makebox(0,0)[cc]{\small 186}}
\put(46.80,53.67){\makebox(0,0)[cc]{\small 160}}
\put(66.80,53.67){\makebox(0,0)[cc]{\small 340}}
\put(66.80,73.67){\makebox(0,0)[cc]{\small 424}}
\put(66.80,93.67){\makebox(0,0)[cc]{\small 164}}
\put(66.80,113.67){\makebox(0,0)[cc]{\small 98}}
\put(66.80,133.67){\makebox(0,0)[cc]{\small 52}}
\put(66.80,153.67){\makebox(0,0)[cc]{\small 16}}
\put(66.80,173.67){\makebox(0,0)[cc]{\small 4}}

\put(3,39.67){\circle*{2.00}}
\put(3,79.67){\circle*{2.00}}
\put(3,59.67){\circle*{2.00}}
\put(3,99.67){\circle*{2.00}}

\put(3,43.67){\makebox(0,0)[cc]{\small 4}}
\put(3,83.67){\makebox(0,0)[cc]{\small 6}}
\put(3,63.67){\makebox(0,0)[cc]{\small 12}}
\put(3,103.34){\makebox(0,0)[cc]{\small 3}}

\put(13,59.67){\circle*{2.00}}
\put(13,119.67){\circle*{2.00}}
\put(13,99.67){\circle*{2.00}}
\put(13,79.67){\circle*{2.00}}

\put(13,83.67){\makebox(0,0)[cc]{\small 13}}
\put(13,63.67){\makebox(0,0)[cc]{\small 16}}
\put(13,103.34){\makebox(0,0)[cc]{\small 4}}
\put(13,123.67){\makebox(0,0)[cc]{\small 1}}

\put(23,119.67){\circle*{2.00}}
\put(23,59.67){\circle*{2.00}}
\put(23,99.67){\circle*{2.00}}
\put(23,79.67){\circle*{2.00}}

\put(23,43.67){\makebox(0,0)[cc]{\small 15}}
\put(23,83.67){\makebox(0,0)[cc]{\small 22}}
\put(23,63.67){\makebox(0,0)[cc]{\small 24}}
\put(23,103.34){\makebox(0,0)[cc]{\small 7}}
\put(23,123.67){\makebox(0,0)[cc]{\small 3}}

\put(33,119.67){\circle*{2.00}}
\put(33,59.67){\circle*{2.00}}
\put(33,139.67){\circle*{2.00}}
\put(33,79.67){\circle*{2.00}}
\put(33,99.67){\circle*{2.00}}

\put(33,83.67){\makebox(0,0)[cc]{\small 31}}
\put(33,63.67){\makebox(0,0)[cc]{\small 41}}
\put(33,103.67){\makebox(0,0)[cc]{\small 16}}
\put(33,123.67){\makebox(0,0)[cc]{\small 4}}
\put(33,143.67){\makebox(0,0)[cc]{\small 1}}

\put(43,119.67){\circle*{2.00}}
\put(43,59.67){\circle*{2.00}}
\put(43,139.67){\circle*{2.00}}
\put(43,79.67){\circle*{2.00}}
\put(43,99.67){\circle*{2.00}}

\put(43,63.67){\makebox(0,0)[cc]{\small 63}}
\put(43,83.67){\makebox(0,0)[cc]{\small 48}}
\put(43,103.34){\makebox(0,0)[cc]{\small 27}}
\put(43,123.67){\makebox(0,0)[cc]{\small 7}}
\put(43,143.67){\makebox(0,0)[cc]{\small 3}}

\put(53,119.67){\circle*{2.00}}
\put(53,59.67){\circle*{2.00}}
\put(53,139.67){\circle*{2.00}}
\put(53,79.67){\circle*{2.00}}
\put(53,159.67){\circle*{2.00}}
\put(53,99.67){\circle*{2.00}}

\put(53,123.67){\makebox(0,0)[cc]{\small 16}}
\put(53,83.67){\makebox(0,0)[cc]{\small 79}}
\put(53,103.34){\makebox(0,0)[cc]{\small 40}}
\put(53,63.67){\makebox(0,0)[cc]{\small 87}}
\put(53,143.67){\makebox(0,0)[cc]{\small 4}}
\put(53,163.67){\makebox(0,0)[cc]{\small 1}}

\put(63,119.67){\circle*{2.00}}
\put(63,59.67){\circle*{2.00}}
\put(63,139.67){\circle*{2.00}}
\put(63,79.67){\circle*{2.00}}
\put(63,159.67){\circle*{2.00}}
\put(63,99.67){\circle*{2.00}}

\put(63,43.67){\makebox(0,0)[cc]{\small 64}}
\put(63,123.67){\makebox(0,0)[cc]{\small 28}}
\put(63,83.67){\makebox(0,0)[cc]{\small 121}}
\put(63,103.67){\makebox(0,0)[cc]{\small 64}}
\put(63,63.67){\makebox(0,0)[cc]{\small 125}}
\put(63,143.67){\makebox(0,0)[cc]{\small 7}}
\put(63,163.67){\makebox(0,0)[cc]{\small 3}}

\put(13,39.67){\circle*{2.00}}
\put(23,39.67){\circle*{2.00}}
\put(33,39.67){\circle*{2.00}}
\put(43,39.67){\circle*{2.00}}
\put(53,39.67){\circle*{2.00}}
\put(63,39.67){\circle*{2.00}}

\put(13,43.67){\makebox(0,0)[cc]{\small 9}}
\put(33,43.67){\makebox(0,0)[cc]{\small 18}}
\put(43,43.67){\makebox(0,0)[cc]{\small 24}}
\put(53,43.67){\makebox(0,0)[cc]{\small 42}}
\end{picture}

\newpage

\begin{picture}(400,530)(-190.00,40.00)
\special{em:linewidth 0.4pt}
\unitlength 1.10mm
\linethickness{0.4pt}
\put(-53.25,27.00){\vector(0,1){150.00}}
\multiput(-53.25,49.67)(0.00,10.00){13}{\line(-1,0){1.00}}
\put(-60.00,49.67){\makebox(0,0)[r]{\small $\frac{1}{2}$}}
\put(-60.00,59.67){\makebox(0,0)[r]{\small $1$}}
\put(-60.00,69.67){\makebox(0,0)[r]{\small $\frac{3}{2}$}}
\put(-60.00,79.67){\makebox(0,0)[r]{\small $2$}}
\put(-60.00,89.67){\makebox(0,0)[r]{\small $\frac{5}{2}$}}
\put(-60.00,99.67){\makebox(0,0)[r]{\small $3$}}
\put(-60.00,109.67){\makebox(0,0)[r]{\small $\frac{7}{2}$}}
\put(-60.00,119.67){\makebox(0,0)[r]{\small $4$}}
\put(-60.00,129.67){\makebox(0,0)[r]{\small $\frac{9}{2}$}}
\put(-60.00,139.67){\makebox(0,0)[r]{\small $5$}}
\put(-60.00,149.67){\makebox(0,0)[r]{\small $\frac{11}{2}$}}
\put(-60.00,159.67){\makebox(0,0)[r]{\small $6$}}
\put(-60.00,169.67){\makebox(0,0)[r]{\small $\frac{13}{2}$}}

\put(-70.00,39.67){\vector(1,0){140.00}}
\put(60,30.67){\makebox(0,0)[cc]{\small $m^2/\alpha$}}
\put(-52.375,175.00){\makebox(0,0)[l]{\small spin}}
\put(3.00,25.00){\makebox(0,0)[cc]{\rm Fig.2}}

\put(-50.00,130.00){\framebox(33,17)[cc]{\shortstack[l]{$D=4$\\
                                            {\circle*{2.00}}
                                            $\alpha (0)_{\rm NS}
                                            = -\frac{3}{16}$\\
                                            {\circle{2.00}}
                                            $\alpha (0)_{\rm R}
                                            = 0$ }}}

\put(-47.00,59.67){\circle*{2.00}}

\put(-53.25,49.67){\circle{2.00}}
\put(-33.25,49.67){\circle{2.00}}
\put(-33.25,69.67){\circle{2.00}}
\put(-13.25,49.67){\circle{2.00}}
\put(-13.25,69.67){\circle{2.00}}
\put(-13.25,89.67){\circle{2.00}}
\put(6.75,49.67){\circle{2.00}}
\put(6.75,69.67){\circle{2.00}}
\put(6.75,89.67){\circle{2.00}}
\put(6.75,109.67){\circle{2.00}}
\put(26.75,49.67){\circle{2.00}}
\put(26.75,69.67){\circle{2.00}}
\put(26.75,89.67){\circle{2.00}}
\put(26.75,109.67){\circle{2.00}}
\put(26.75,129.67){\circle{2.00}}
\put(46.75,49.67){\circle{2.00}}
\put(46.75,69.67){\circle{2.00}}
\put(46.75,89.67){\circle{2.00}}
\put(46.75,109.67){\circle{2.00}}
\put(46.75,129.67){\circle{2.00}}
\put(46.75,149.67){\circle{2.00}}
\put(66.75,49.67){\circle{2.00}}
\put(66.75,69.67){\circle{2.00}}
\put(66.75,89.67){\circle{2.00}}
\put(66.75,109.67){\circle{2.00}}
\put(66.75,129.67){\circle{2.00}}
\put(66.75,149.67){\circle{2.00}}
\put(66.75,169.67){\circle{2.00}}

\put(-27.00,39.67){\circle*{2.00}}
\put(-27.00,59.67){\circle*{2.00}}
\put(-27.00,79.67){\circle*{2.00}}


\put(-7.00,39.67){\circle*{2.00}}
\put(-7.00,79.67){\circle*{2.00}}
\put(-7.00,59.67){\circle*{2.00}}
\put(-7.00,99.67){\circle*{2.00}}

\put(-47,63.67){\makebox(0,0)[cc]{\small 1}}
\put(-27,83.67){\makebox(0,0)[cc]{\small 1}}
\put(-27,63.67){\makebox(0,0)[cc]{\small 2}}
\put(-27,43.67){\makebox(0,0)[cc]{\small 2}}


\put(-7,103.67){\makebox(0,0)[cc]{\small 1}}
\put(-7,63.67){\makebox(0,0)[cc]{\small 7}}
\put(-7,83.67){\makebox(0,0)[cc]{\small 4}}
\put(-7,43.67){\makebox(0,0)[cc]{\small 3}}

\put(-51.20,53.67){\makebox(0,0)[cc]{\small 1}}
\put(-33.20,73.67){\makebox(0,0)[cc]{\small 2}}
\put(-33.20,53.67){\makebox(0,0)[cc]{\small 2}}
\put(-13.20,93.67){\makebox(0,0)[cc]{\small 2}}
\put(-13.20,73.67){\makebox(0,0)[cc]{\small 6}}
\put(-13.20,53.67){\makebox(0,0)[cc]{\small 6}}
\put(6.80,113.67){\makebox(0,0)[cc]{\small 2}}
\put(6.80,93.67){\makebox(0,0)[cc]{\small 8}}
\put(6.80,73.67){\makebox(0,0)[cc]{\small 16}}
\put(6.80,53.67){\makebox(0,0)[cc]{\small 16}}
\put(26.80,133.67){\makebox(0,0)[cc]{\small 2}}
\put(26.80,113.67){\makebox(0,0)[cc]{\small 8}}
\put(26.80,93.67){\makebox(0,0)[cc]{\small 24}}
\put(26.80,73.67){\makebox(0,0)[cc]{\small 42}}
\put(26.80,53.67){\makebox(0,0)[cc]{\small 36}}
\put(46.80,153.67){\makebox(0,0)[cc]{\small 2}}
\put(46.80,133.67){\makebox(0,0)[cc]{\small 8}}
\put(46.80,113.67){\makebox(0,0)[cc]{\small 26}}
\put(46.80,93.67){\makebox(0,0)[cc]{\small 64}}
\put(46.80,73.67){\makebox(0,0)[cc]{\small 98}}
\put(46.80,53.67){\makebox(0,0)[cc]{\small 80}}
\put(66.80,53.67){\makebox(0,0)[cc]{\small 170}}
\put(66.80,73.67){\makebox(0,0)[cc]{\small 212}}
\put(66.80,93.67){\makebox(0,0)[cc]{\small 132}}
\put(66.80,113.67){\makebox(0,0)[cc]{\small 49}}
\put(66.80,133.67){\makebox(0,0)[cc]{\small 26}}
\put(66.80,153.67){\makebox(0,0)[cc]{\small 8}}
\put(66.80,173.67){\makebox(0,0)[cc]{\small 2}}

\put(13,59.67){\circle*{2.00}}
\put(13,119.67){\circle*{2.00}}
\put(13,99.67){\circle*{2.00}}
\put(13,79.67){\circle*{2.00}}

\put(13,83.67){\makebox(0,0)[cc]{\small 13}}
\put(13,63.67){\makebox(0,0)[cc]{\small 16}}
\put(13,103.34){\makebox(0,0)[cc]{\small 4}}
\put(13,123.67){\makebox(0,0)[cc]{\small 1}}

\put(33,119.67){\circle*{2.00}}
\put(33,59.67){\circle*{2.00}}
\put(33,139.67){\circle*{2.00}}
\put(33,79.67){\circle*{2.00}}
\put(33,99.67){\circle*{2.00}}

\put(33,83.67){\makebox(0,0)[cc]{\small 31}}
\put(33,63.67){\makebox(0,0)[cc]{\small 41}}
\put(33,103.67){\makebox(0,0)[cc]{\small 16}}
\put(33,123.67){\makebox(0,0)[cc]{\small 4}}
\put(33,143.67){\makebox(0,0)[cc]{\small 1}}

\put(53,119.67){\circle*{2.00}}
\put(53,59.67){\circle*{2.00}}
\put(53,139.67){\circle*{2.00}}
\put(53,79.67){\circle*{2.00}}
\put(53,159.67){\circle*{2.00}}
\put(53,99.67){\circle*{2.00}}

\put(53,123.67){\makebox(0,0)[cc]{\small 16}}
\put(53,83.67){\makebox(0,0)[cc]{\small 79}}
\put(53,103.34){\makebox(0,0)[cc]{\small 40}}
\put(53,63.67){\makebox(0,0)[cc]{\small 87}}
\put(53,143.67){\makebox(0,0)[cc]{\small 4}}
\put(53,163.67){\makebox(0,0)[cc]{\small 1}}

\put(13,39.67){\circle*{2.00}}
\put(33,39.67){\circle*{2.00}}
\put(53,39.67){\circle*{2.00}}

\put(13,43.67){\makebox(0,0)[cc]{\small 9}}
\put(33,43.67){\makebox(0,0)[cc]{\small 18}}
\put(53,43.67){\makebox(0,0)[cc]{\small 42}}
\end{picture}

\newpage
\begin{picture}(400,530)(-190.00,40.00)
\special{em:linewidth 0.4pt}
\unitlength 1.10mm
\linethickness{0.4pt}
\put(-55.125,22.00){\vector(0,1){140.00}}
\multiput(-55.125,39.67)(0.00,10.00){10}{\line(-1,0){1.00}}
\put(-61.125,39.67){\makebox(0,0)[r]{\small $\frac{1}{2}$}}
\put(-61.125,59.67){\makebox(0,0)[r]{\small $\frac{3}{2}$}}
\put(-61.125,79.67){\makebox(0,0)[r]{\small $\frac{5}{2}$}}
\put(-61.125,99.67){\makebox(0,0)[r]{\small $\frac{7}{2}$}}
\put(-61.125,119.67){\makebox(0,0)[r]{\small $\frac{9}{2}$}}
\put(-61.125,49.67){\makebox(0,0)[r]{\small $1$}}
\put(-61.125,69.67){\makebox(0,0)[r]{\small $2$}}
\put(-61.125,89.67){\makebox(0,0)[r]{\small $3$}}
\put(-61.125,109.67){\makebox(0,0)[r]{\small $4$}}
\put(-61.125,129.67){\makebox(0,0)[r]{\small $5$}}

\put(-64.50,29.67){\vector(1,0){120.00}}
\put(45,20.67){\makebox(0,0)[cc]{\small $m^2/4\alpha$}}
\put(-63.375,158.00){\makebox(0,0)[l]{\small spin}}
\put(4.00,15.00){\makebox(0,0)[cc]{\rm Fig.3}}

\put(-50.00,140.00){\framebox(37,17)[cc]{\shortstack[l]{$D=4$\\
                                           {\circle*{2.00}}
                                           $\alpha (0)_{\rm NS-NS}
                                           = -\frac{6}{16}$ \\
                                           {\circle{2.00}}
                                           $\alpha (0)_{\rm R-R}
                                           = 0$ }}}

\put(-58.875,29.67){\circle*{2.00}}
\put(-38.875,69.67){\circle*{2.00}}
\put(-38.875,29.67){\circle*{2.00}}
\put(-38.875,49.67){\circle*{2.00}}
\put(21.125,129.67){\circle*{2.00}}
\put(21.125,29.67){\circle*{2.00}}
\put(21.125,89.67){\circle*{2.00}}
\put(21.125,49.67){\circle*{2.00}}
\put(21.125,109.67){\circle*{2.00}}
\put(21.125,69.67){\circle*{2.00}}
\put(-18.875,29.67){\circle*{2.00}}
\put(-18.875,89.67){\circle*{2.00}}
\put(-18.875,49.67){\circle*{2.00}}
\put(-18.875,109.67){\circle*{2.00}}
\put(-18.875,69.67){\circle*{2.00}}
\put(41.125,29.67){\circle*{2.00}}
\put(41.125,89.67){\circle*{2.00}}
\put(41.125,49.67){\circle*{2.00}}
\put(41.125,109.67){\circle*{2.00}}
\put(41.125,69.67){\circle*{2.00}}
\put(41.125,69.67){\circle*{2.00}}
\put(41.125,129.67){\circle*{2.00}}
\put(1.125,29.67){\circle*{2.00}}
\put(1.125,89.67){\circle*{2.00}}
\put(1.125,49.67){\circle*{2.00}}
\put(1.125,109.67){\circle*{2.00}}
\put(1.125,69.67){\circle*{2.00}}
\put(1.125,129.67){\circle*{2.00}}
\put(-48.875,69.67){\circle*{2.00}}
\put(-48.875,29.67){\circle*{2.00}}
\put(-48.875,49.67){\circle*{2.00}}
\put(11.125,129.67){\circle*{2.00}}
\put(11.125,29.67){\circle*{2.00}}
\put(11.125,89.67){\circle*{2.00}}
\put(11.125,49.67){\circle*{2.00}}
\put(11.125,109.67){\circle*{2.00}}
\put(11.125,69.67){\circle*{2.00}}
\put(-28.875,29.67){\circle*{2.00}}
\put(-28.875,89.67){\circle*{2.00}}
\put(-28.875,49.67){\circle*{2.00}}
\put(-28.875,109.67){\circle*{2.00}}
\put(-28.875,69.67){\circle*{2.00}}
\put(31.125,29.67){\circle*{2.00}}
\put(31.125,89.67){\circle*{2.00}}
\put(31.125,49.67){\circle*{2.00}}
\put(31.125,109.67){\circle*{2.00}}
\put(31.125,69.67){\circle*{2.00}}
\put(31.125,69.67){\circle*{2.00}}
\put(31.125,129.67){\circle*{2.00}}
\put(-8.875,29.67){\circle*{2.00}}
\put(-8.875,89.67){\circle*{2.00}}
\put(-8.875,49.67){\circle*{2.00}}
\put(-8.875,109.67){\circle*{2.00}}
\put(-8.875,69.67){\circle*{2.00}}
\put(-8.875,129.67){\circle*{2.00}}

\put(-55.125,29.67){\circle{2.00}}
\put(-55.125,49.67){\circle{2.00}}
\put(-35.125,29.67){\circle{2.00}}
\put(-35.125,49.67){\circle{2.00}}
\put(-35.125,69.67){\circle{2.00}}
\put(-35.125,89.67){\circle{2.00}}
\put(-15.125,29.67){\circle{2.00}}
\put(-15.125,49.67){\circle{2.00}}
\put(-15.125,69.67){\circle{2.00}}
\put(-15.125,89.67){\circle{2.00}}
\put(-15.125,109.67){\circle{2.00}}
\put(-15.125,129.67){\circle{2.00}}
\put(4.875,29.67){\circle{2.00}}
\put(4.875,49.67){\circle{2.00}}
\put(4.875,69.67){\circle{2.00}}
\put(4.875,89.67){\circle{2.00}}
\put(4.875,109.67){\circle{2.00}}
\put(4.875,129.67){\circle{2.00}}
\put(24.875,29.67){\circle{2.00}}
\put(24.875,49.67){\circle{2.00}}
\put(24.875,69.67){\circle{2.00}}
\put(24.875,89.67){\circle{2.00}}
\put(24.875,109.67){\circle{2.00}}
\put(24.875,129.67){\circle{2.00}}
\put(44.875,29.67){\circle{2.00}}
\put(44.875,49.67){\circle{2.00}}
\put(44.875,69.67){\circle{2.00}}
\put(44.875,89.67){\circle{2.00}}
\put(44.875,109.67){\circle{2.00}}
\put(44.875,129.67){\circle{2.00}}

\put(-38.875,32.67){\makebox(0,0)[cc]{\small 4}}
\put(-38.875,52.67){\makebox(0,0)[cc]{\small 4}}
\put(-38.875,72.67){\makebox(0,0)[cc]{\small 4}}
\put(-58.875,32.67){\makebox(0,0)[cc]{\small 1}}
\put(-48.875,32.67){\makebox(0,0)[cc]{\small 1}}
\put(-48.875,52.67){\makebox(0,0)[cc]{\small 1}}
\put(-48.875,72.67){\makebox(0,0)[cc]{\small 1}}
\put(-28.875,32.67){\makebox(0,0)[cc]{\small 9}}
\put(-28.875,52.67){\makebox(0,0)[cc]{\small 17}}
\put(-28.875,72.67){\makebox(0,0)[cc]{\small 13}}
\put(-28.875,92.67){\makebox(0,0)[cc]{\small 5}}
\put(-28.875,112.67){\makebox(0,0)[cc]{\small 1}}
\put(-18.875,32.67){\makebox(0,0)[cc]{\small 27}}
\put(-18.875,52.67){\makebox(0,0)[cc]{\small 54}}
\put(-18.875,72.67){\makebox(0,0)[cc]{\small 54}}
\put(-18.875,92.67){\makebox(0,0)[cc]{\small 27}}
\put(-18.875,112.67){\makebox(0,0)[cc]{\small 9}}
\put(1.125,32.67){\makebox(0,0)[cc]{\small 205}}
\put(1.125,52.67){\makebox(0,0)[cc]{\small 124}}
\put(1.125,72.67){\makebox(0,0)[cc]{\small 489}}
\put(1.125,92.67){\makebox(0,0)[cc]{\small 321}}
\put(1.125,112.67){\makebox(0,0)[cc]{\small 153}}
\put(1.125,132.67){\makebox(0,0)[cc]{\small 45}}
\put(11.125,32.67){\makebox(0,0)[cc]{\small 523}}
\put(11.125,52.67){\makebox(0,0)[cc]{\small 1285}}
\put(11.125,72.67){\makebox(0,0)[cc]{\small 1358}}
\put(11.125,92.67){\makebox(0,0)[cc]{\small 972}}
\put(11.125,112.67){\makebox(0,0)[cc]{\small 502}}
\put(11.125,132.67){\makebox(0,0)[cc]{\small 187}}
\put(21.125,32.67){\makebox(0,0)[cc]{\small 4343}}
\put(21.125,52.67){\makebox(0,0)[cc]{\small 3244}}
\put(21.125,72.67){\makebox(0,0)[cc]{\small 3652}}
\put(21.125,92.67){\makebox(0,0)[cc]{\small 2770}}
\put(21.125,112.67){\makebox(0,0)[cc]{\small 1594}}
\put(21.125,132.67){\makebox(0,0)[cc]{\small 684}}
\put(31.125,32.67){\makebox(0,0)[cc]{\small 7239}}
\put(31.125,52.67){\makebox(0,0)[cc]{\small 8061}}
\put(31.125,72.67){\makebox(0,0)[cc]{\small 9293}}
\put(31.125,92.67){\makebox(0,0)[cc]{\small 7462}}
\put(31.125,112.67){\makebox(0,0)[cc]{\small 4570}}
\put(31.125,132.67){\makebox(0,0)[cc]{\small 2189}}
\put(41.125,32.67){\makebox(0,0)[cc]{\small 7636}}
\put(41.125,52.67){\makebox(0,0)[cc]{\small 19144}}
\put(41.125,72.67){\makebox(0,0)[cc]{\small 22660}}
\put(41.125,92.67){\makebox(0,0)[cc]{\small 18853}}
\put(41.125,112.67){\makebox(0,0)[cc]{\small 12223}}
\put(41.125,132.67){\makebox(0,0)[cc]{\small 6325}}
\put(-8.875,32.67){\makebox(0,0)[cc]{\small 75}}
\put(-8.875,52.67){\makebox(0,0)[cc]{\small 172}}
\put(-8.875,72.67){\makebox(0,0)[cc]{\small 168}}
\put(-8.875,92.67){\makebox(0,0)[cc]{\small 101}}
\put(-8.875,112.67){\makebox(0,0)[cc]{\small 39}}
\put(-8.875,132.67){\makebox(0,0)[cc]{\small 9}}

\put(-53.150,26.67){\makebox(0,0)[cc]{\small 4}}
\put(-53.150,46.67){\makebox(0,0)[cc]{\small 4}}
\put(-35.150,26.67){\makebox(0,0)[cc]{\small 32}}
\put(-35.150,46.67){\makebox(0,0)[cc]{\small 64}}
\put(-35.150,66.67){\makebox(0,0)[cc]{\small 48}}
\put(-35.150,86.67){\makebox(0,0)[cc]{\small 16}}
\put(-15.150,26.67){\makebox(0,0)[cc]{\small 304}}
\put(-15.150,46.67){\makebox(0,0)[cc]{\small 688}}
\put(-15.150,66.67){\makebox(0,0)[cc]{\small 640}}
\put(-15.150,86.67){\makebox(0,0)[cc]{\small 342}}
\put(-15.150,106.67){\makebox(0,0)[cc]{\small 112}}
\put(-15.150,126.67){\makebox(0,0)[cc]{\small 16}}
\put(4.850,26.67){\makebox(0,0)[cc]{\small 2320}}
\put(4.850,46.67){\makebox(0,0)[cc]{\small 5520}}
\put(4.850,66.67){\makebox(0,0)[cc]{\small 5776}}
\put(4.850,86.67){\makebox(0,0)[cc]{\small 3584}}
\put(4.850,106.67){\makebox(0,0)[cc]{\small 1936}}
\put(4.850,126.67){\makebox(0,0)[cc]{\small 656}}
\put(24.850,26.67){\makebox(0,0)[cc]{\small 14816}}
\put(24.850,46.67){\makebox(0,0)[cc]{\small 36640}}
\put(24.850,66.67){\makebox(0,0)[cc]{\small 41440}}
\put(24.850,86.67){\makebox(0,0)[cc]{\small 32320}}
\put(24.850,106.67){\makebox(0,0)[cc]{\small 18928}}
\put(24.850,126.67){\makebox(0,0)[cc]{\small 8480}}
\put(44.850,26.67){\makebox(0,0)[cc]{\small 83376}}
\put(44.850,46.67){\makebox(0,0)[cc]{\small 211376}}
\put(44.850,66.67){\makebox(0,0)[cc]{\small 251632}}
\put(44.850,86.67){\makebox(0,0)[cc]{\small 212848}}
\put(44.850,106.67){\makebox(0,0)[cc]{\small 140160}}
\put(44.850,126.67){\makebox(0,0)[cc]{\small 74608}}
\put(-8.875,32.67){\makebox(0,0)[cc]{\small 75}}
\put(-8.875,52.67){\makebox(0,0)[cc]{\small 172}}
\put(-8.875,72.67){\makebox(0,0)[cc]{\small 168}}
\put(-8.875,92.67){\makebox(0,0)[cc]{\small 101}}
\put(-8.875,112.67){\makebox(0,0)[cc]{\small 39}}
\put(-8.875,132.67){\makebox(0,0)[cc]{\small 9}}

\end{picture}

\newpage

\begin{picture}(400,530)(-190.00,40.00)
\special{em:linewidth 0.4pt}
\unitlength 1.10mm
\linethickness{0.4pt}
\put(-55.125,22.00){\vector(0,1){140.00}}
\multiput(-55.125,39.67)(0.00,10.00){10}{\line(-1,0){1.00}}
\put(-61.125,39.67){\makebox(0,0)[r]{\small $\frac{1}{2}$}}
\put(-61.125,59.67){\makebox(0,0)[r]{\small $\frac{3}{2}$}}
\put(-61.125,79.67){\makebox(0,0)[r]{\small $\frac{5}{2}$}}
\put(-61.125,99.67){\makebox(0,0)[r]{\small $\frac{7}{2}$}}
\put(-61.125,119.67){\makebox(0,0)[r]{\small $\frac{9}{2}$}}
\put(-61.125,49.67){\makebox(0,0)[r]{\small $1$}}
\put(-61.125,69.67){\makebox(0,0)[r]{\small $2$}}
\put(-61.125,89.67){\makebox(0,0)[r]{\small $3$}}
\put(-61.125,109.67){\makebox(0,0)[r]{\small $4$}}
\put(-61.125,129.67){\makebox(0,0)[r]{\small $5$}}

\put(-64.50,29.67){\vector(1,0){120.00}}
\put(45,20.67){\makebox(0,0)[cc]{\small $m^2/4\alpha$}}
\put(-63.375,158.00){\makebox(0,0)[l]{\small spin}}
\put(4.00,15.00){\makebox(0,0)[cc]{\rm Fig.4}}

\put(-50.00,140.00){\framebox(37,17)[cc]{\shortstack[l]{$D=4$\\
                                           {\circle*{2.00}}
                                           $\alpha (0)_{\rm NS-NS}
                                           = -\frac{6}{16}$ \\
                                           {\circle{2.00}}
                                           $\alpha (0)_{\rm R-R}
                                           = 0$ }}}

\put(-48.875,69.67){\circle*{2.00}}
\put(-48.875,29.67){\circle*{2.00}}
\put(-48.875,49.67){\circle*{2.00}}
\put(11.125,129.67){\circle*{2.00}}
\put(11.125,29.67){\circle*{2.00}}
\put(11.125,89.67){\circle*{2.00}}
\put(11.125,49.67){\circle*{2.00}}
\put(11.125,109.67){\circle*{2.00}}
\put(11.125,69.67){\circle*{2.00}}
\put(-28.875,29.67){\circle*{2.00}}
\put(-28.875,89.67){\circle*{2.00}}
\put(-28.875,49.67){\circle*{2.00}}
\put(-28.875,109.67){\circle*{2.00}}
\put(-28.875,69.67){\circle*{2.00}}
\put(31.125,29.67){\circle*{2.00}}
\put(31.125,89.67){\circle*{2.00}}
\put(31.125,49.67){\circle*{2.00}}
\put(31.125,109.67){\circle*{2.00}}
\put(31.125,69.67){\circle*{2.00}}
\put(31.125,69.67){\circle*{2.00}}
\put(31.125,129.67){\circle*{2.00}}
\put(-8.875,29.67){\circle*{2.00}}
\put(-8.875,89.67){\circle*{2.00}}
\put(-8.875,49.67){\circle*{2.00}}
\put(-8.875,109.67){\circle*{2.00}}
\put(-8.875,69.67){\circle*{2.00}}
\put(-8.875,129.67){\circle*{2.00}}

\put(-55.125,29.67){\circle{2.00}}
\put(-55.125,49.67){\circle{2.00}}
\put(-35.125,29.67){\circle{2.00}}
\put(-35.125,49.67){\circle{2.00}}
\put(-35.125,69.67){\circle{2.00}}
\put(-35.125,89.67){\circle{2.00}}
\put(-15.125,29.67){\circle{2.00}}
\put(-15.125,49.67){\circle{2.00}}
\put(-15.125,69.67){\circle{2.00}}
\put(-15.125,89.67){\circle{2.00}}
\put(-15.125,109.67){\circle{2.00}}
\put(-15.125,129.67){\circle{2.00}}
\put(4.875,29.67){\circle{2.00}}
\put(4.875,49.67){\circle{2.00}}
\put(4.875,69.67){\circle{2.00}}
\put(4.875,89.67){\circle{2.00}}
\put(4.875,109.67){\circle{2.00}}
\put(4.875,129.67){\circle{2.00}}
\put(24.875,29.67){\circle{2.00}}
\put(24.875,49.67){\circle{2.00}}
\put(24.875,69.67){\circle{2.00}}
\put(24.875,89.67){\circle{2.00}}
\put(24.875,109.67){\circle{2.00}}
\put(24.875,129.67){\circle{2.00}}
\put(44.875,29.67){\circle{2.00}}
\put(44.875,49.67){\circle{2.00}}
\put(44.875,69.67){\circle{2.00}}
\put(44.875,89.67){\circle{2.00}}
\put(44.875,109.67){\circle{2.00}}
\put(44.875,129.67){\circle{2.00}}

\put(-48.875,32.67){\makebox(0,0)[cc]{\small 1}}
\put(-48.875,52.67){\makebox(0,0)[cc]{\small 1}}
\put(-48.875,72.67){\makebox(0,0)[cc]{\small 1}}
\put(-28.875,32.67){\makebox(0,0)[cc]{\small 9}}
\put(-28.875,52.67){\makebox(0,0)[cc]{\small 17}}
\put(-28.875,72.67){\makebox(0,0)[cc]{\small 13}}
\put(-28.875,92.67){\makebox(0,0)[cc]{\small 5}}
\put(-28.875,112.67){\makebox(0,0)[cc]{\small 1}}
\put(11.125,32.67){\makebox(0,0)[cc]{\small 523}}
\put(11.125,52.67){\makebox(0,0)[cc]{\small 1285}}
\put(11.125,72.67){\makebox(0,0)[cc]{\small 1358}}
\put(11.125,92.67){\makebox(0,0)[cc]{\small 972}}
\put(11.125,112.67){\makebox(0,0)[cc]{\small 502}}
\put(11.125,132.67){\makebox(0,0)[cc]{\small 187}}
\put(31.125,32.67){\makebox(0,0)[cc]{\small 7239}}
\put(31.125,52.67){\makebox(0,0)[cc]{\small 8061}}
\put(31.125,72.67){\makebox(0,0)[cc]{\small 9293}}
\put(31.125,92.67){\makebox(0,0)[cc]{\small 7462}}
\put(31.125,112.67){\makebox(0,0)[cc]{\small 4570}}
\put(31.125,132.67){\makebox(0,0)[cc]{\small 2189}}
\put(-8.875,32.67){\makebox(0,0)[cc]{\small 75}}
\put(-8.875,52.67){\makebox(0,0)[cc]{\small 172}}
\put(-8.875,72.67){\makebox(0,0)[cc]{\small 168}}
\put(-8.875,92.67){\makebox(0,0)[cc]{\small 101}}
\put(-8.875,112.67){\makebox(0,0)[cc]{\small 39}}
\put(-8.875,132.67){\makebox(0,0)[cc]{\small 9}}

\put(-53.150,26.67){\makebox(0,0)[cc]{\small 1}}
\put(-53.150,46.67){\makebox(0,0)[cc]{\small 1}}
\put(-35.150,26.67){\makebox(0,0)[cc]{\small 8}}
\put(-35.150,46.67){\makebox(0,0)[cc]{\small 16}}
\put(-35.150,66.67){\makebox(0,0)[cc]{\small 12}}
\put(-35.150,86.67){\makebox(0,0)[cc]{\small 4}}
\put(-15.150,26.67){\makebox(0,0)[cc]{\small 76}}
\put(-15.150,46.67){\makebox(0,0)[cc]{\small 172}}
\put(-15.150,66.67){\makebox(0,0)[cc]{\small 180}}
\put(-15.150,86.67){\makebox(0,0)[cc]{\small 88}}
\put(-15.150,106.67){\makebox(0,0)[cc]{\small 28}}
\put(-15.150,126.67){\makebox(0,0)[cc]{\small 4}}
\put(4.850,26.67){\makebox(0,0)[cc]{\small 580}}
\put(4.850,46.67){\makebox(0,0)[cc]{\small 1380}}
\put(4.850,66.67){\makebox(0,0)[cc]{\small 1444}}
\put(4.850,86.67){\makebox(0,0)[cc]{\small 996}}
\put(4.850,106.67){\makebox(0,0)[cc]{\small 484}}
\put(4.850,126.67){\makebox(0,0)[cc]{\small 164}}
\put(24.850,26.67){\makebox(0,0)[cc]{\small 3704}}
\put(24.850,46.67){\makebox(0,0)[cc]{\small 9160}}
\put(24.850,66.67){\makebox(0,0)[cc]{\small 10360}}
\put(24.850,86.67){\makebox(0,0)[cc]{\small 8080}}
\put(24.850,106.67){\makebox(0,0)[cc]{\small 4732}}
\put(24.850,126.67){\makebox(0,0)[cc]{\small 2140}}
\put(44.850,26.67){\makebox(0,0)[cc]{\small 20844}}
\put(44.850,46.67){\makebox(0,0)[cc]{\small 52844}}
\put(44.850,66.67){\makebox(0,0)[cc]{\small 62908}}
\put(44.850,86.67){\makebox(0,0)[cc]{\small 53212}}
\put(44.850,106.67){\makebox(0,0)[cc]{\small 35040}}
\put(44.850,126.67){\makebox(0,0)[cc]{\small 18656}}
\put(-8.875,32.67){\makebox(0,0)[cc]{\small 75}}
\put(-8.875,52.67){\makebox(0,0)[cc]{\small 172}}
\put(-8.875,72.67){\makebox(0,0)[cc]{\small 168}}
\put(-8.875,92.67){\makebox(0,0)[cc]{\small 101}}
\put(-8.875,112.67){\makebox(0,0)[cc]{\small 39}}
\put(-8.875,132.67){\makebox(0,0)[cc]{\small 9}}

\end{picture}

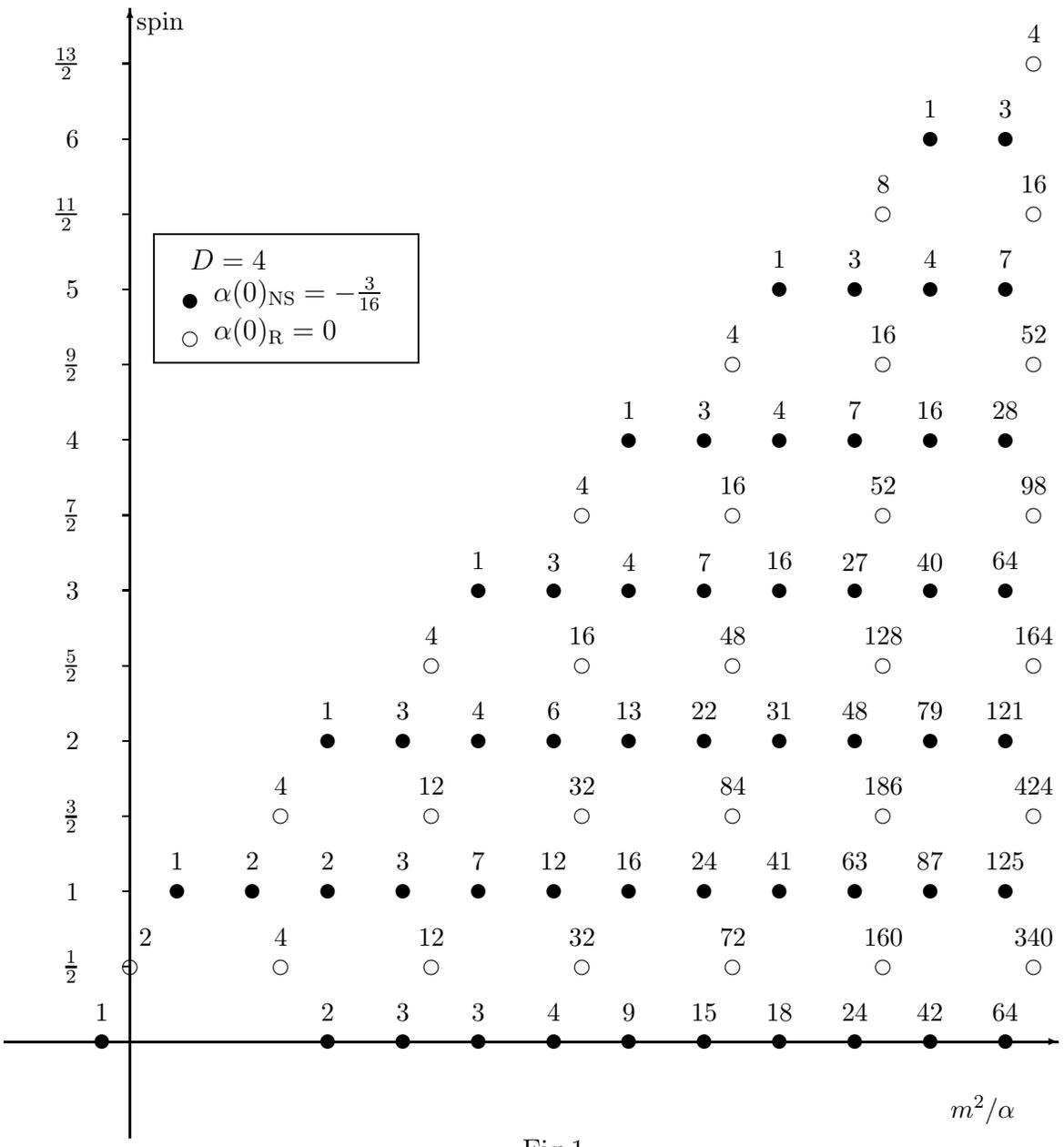
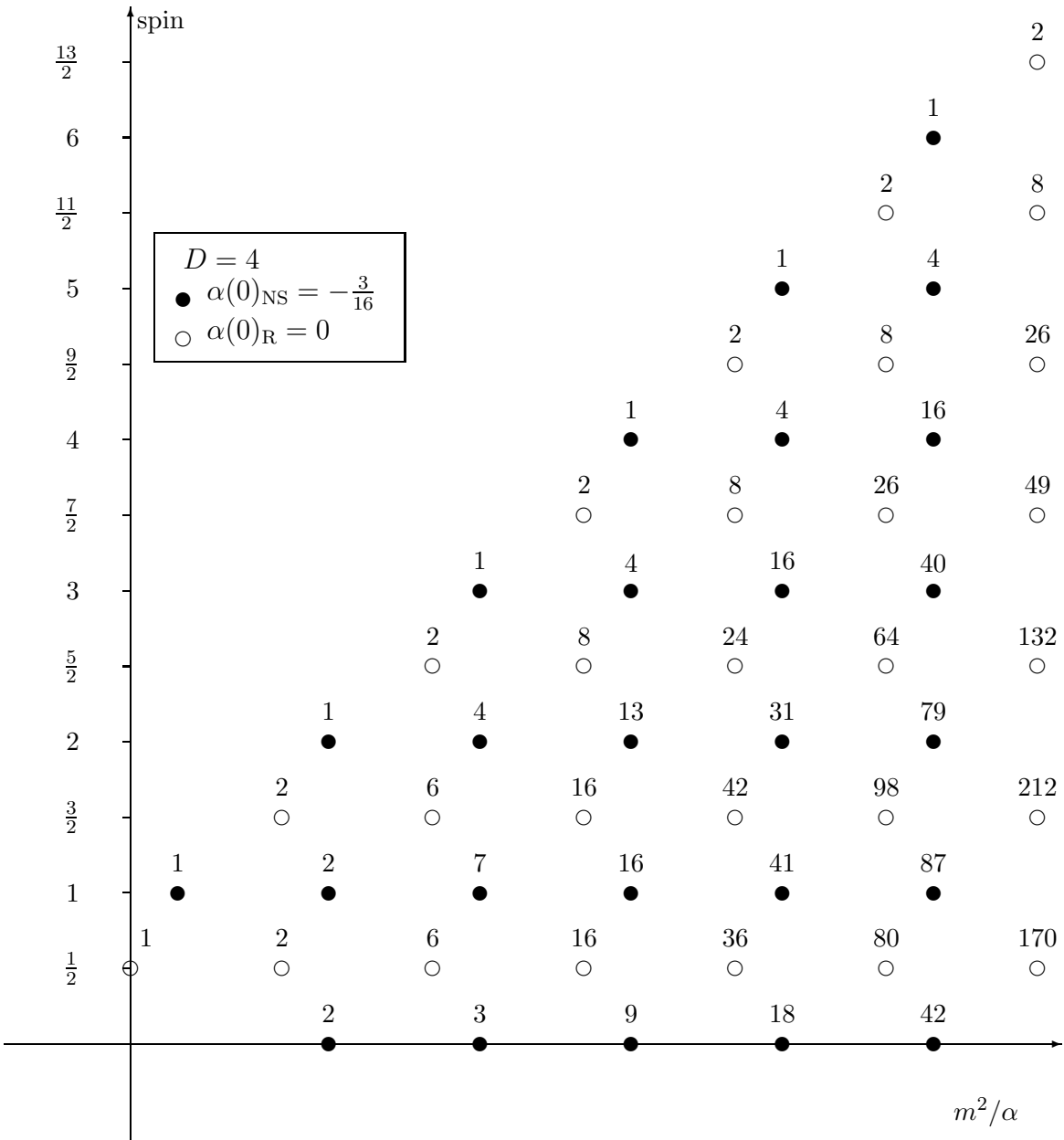
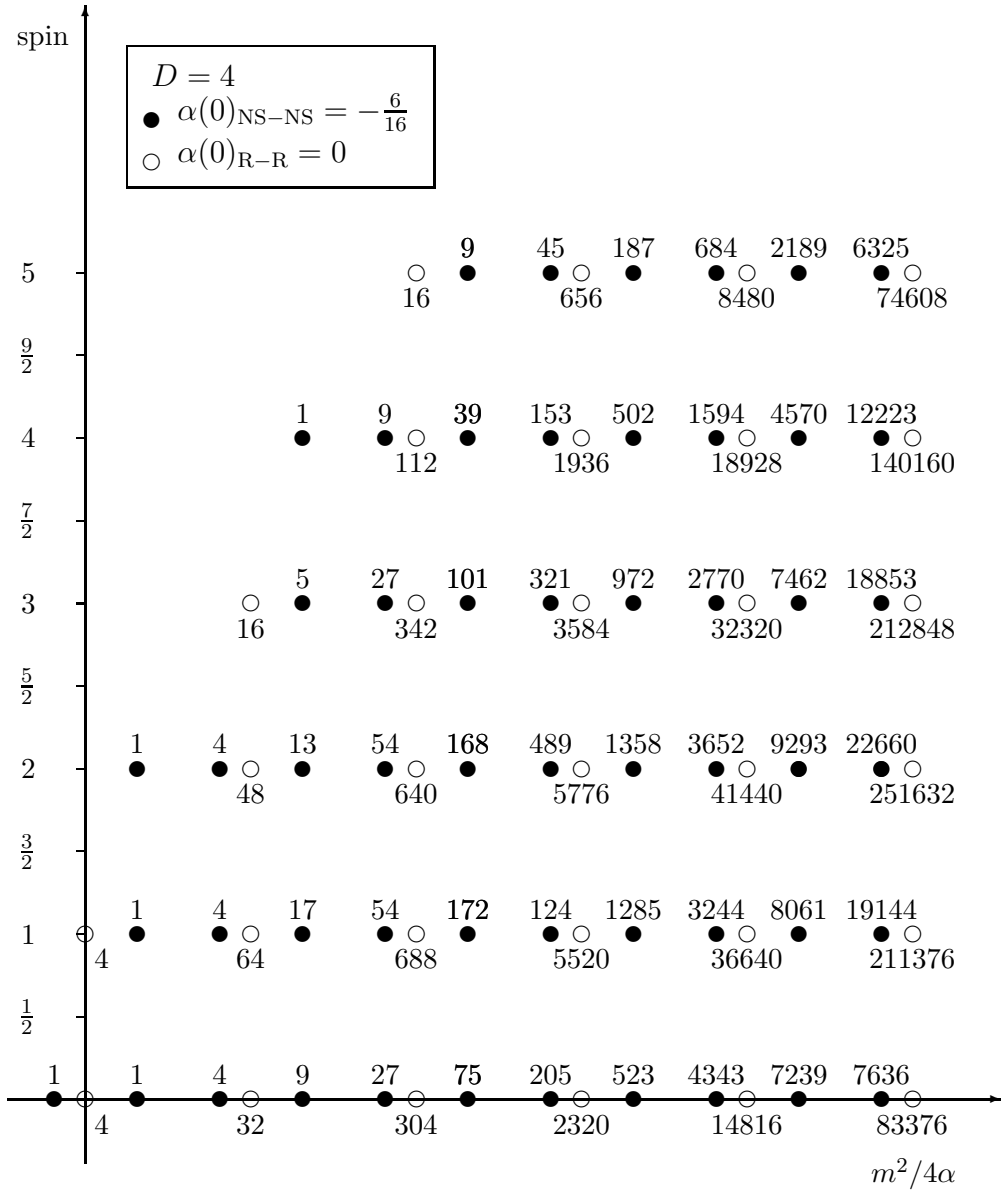
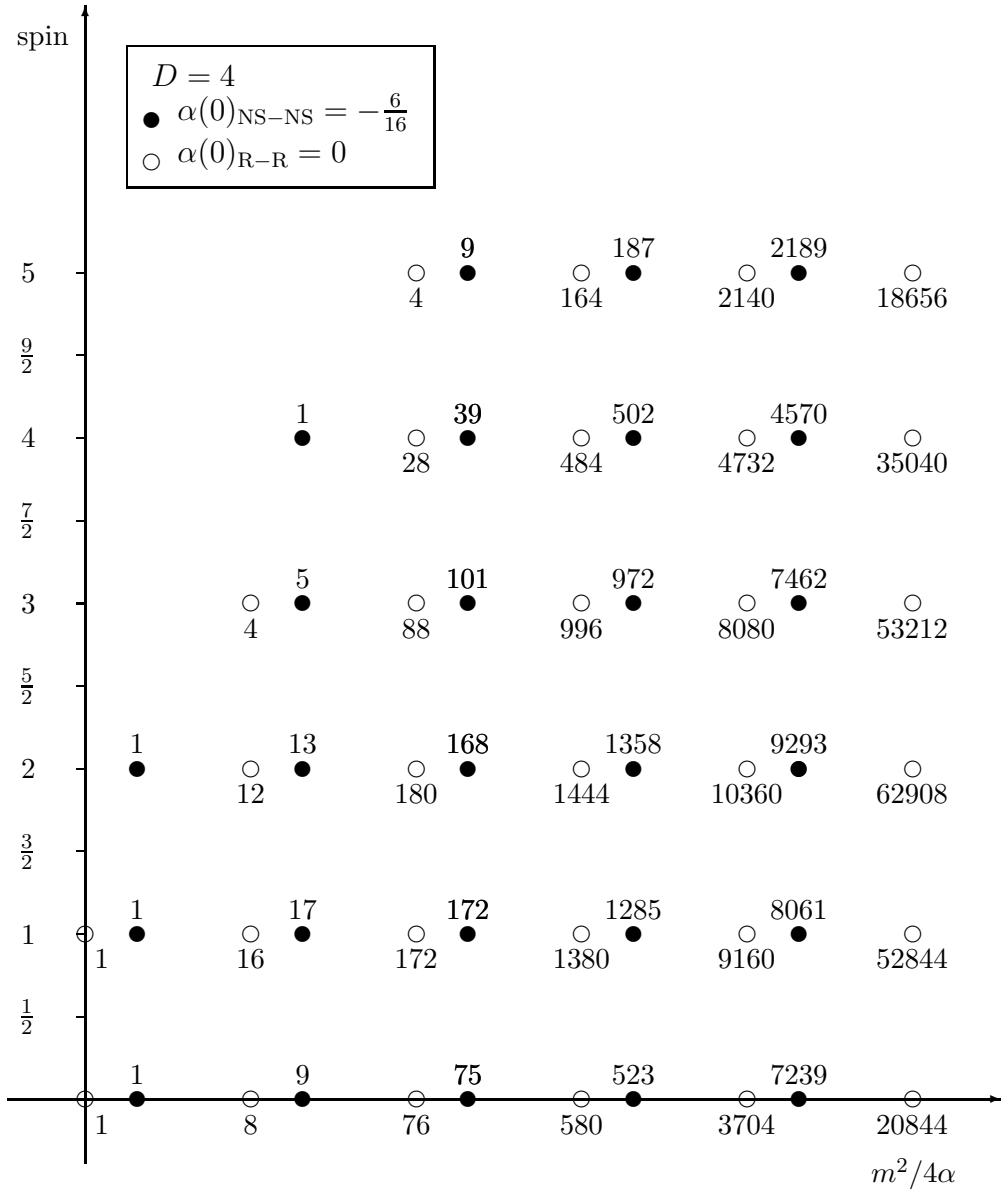
\end{document}